\documentclass{PoS}
\usepackage{wrapfig,rotating}
\usepackage{epsf,epsfig,epstopdf}

\title{Soft and Coulomb gluon resummation in squark-antisquark production at the LHC}

\ShortTitle{Soft and Coulomb gluon resummation in squark-antisquark production at the LHC}

\author{M.~Beneke\\
        Institut f\"ur Theoretische Physik E, 
        RWTH Aachen University,\\  D - 52056 Aachen, Germany
        }

\author{\speaker{P.~Falgari} \thanks{Preprint TTK-10-12, SFB/CPP-10-14, IPPP/10/04, DCPT/10/08, FR-PHENO-2010-007. The work of M.B. is
supported by the DFG Sonder\-forschungs\-be\-reich/Trans\-regio~9
``Computergest\"utzte Theoretische Teilchenphysik''.}\\
        IPPP, Department of Physics, University of Durham, \\
        Durham DH1 3LE, England\\
        E-mail: \email{pietro.falgari@durham.ac.uk}
        }

\author{C.~Schwinn\\
        Albert-Ludwigs Universit\"at Freiburg,\\
        Physikalisches Institut, D-79104 Freiburg, Germany
        }

\abstract{We present results for a combined resummation of
soft and Coloumb gluon corrections in squark-antisquark production
at the LHC, including the non-trivial interference of the two effects and
the contribution of squark-antisquark bound states below threshold.   
The total correction to the next-to-leading order approximation is found to be sizeable, and amounts to $5-14 \%$
in the squark mass region $200 \, \mbox{GeV} -2 \, \mbox{TeV}$. The scale
dependence of the total cross section is also reduced.}

\FullConference{RADCOR 2009 - 9th International Symposium on Radiative Corrections (Applications of Quantum Field Theory to Phenomenology) \\
		 October 25-30 2009\\
		 Ascona, Switzerland}

\begin{document}

\section{Introduction}

The total cross sections for the processes 
$
pp' \rightarrow H H'+X
$, 
 where $pp' \in \{qq,q\bar{q},gg,gq,g
\bar{q}\}$ and $H, \, H'$ are heavy particles in arbitrary representations $R, \, R'$ of the $SU(3)$ colour algebra (top quarks, 
squarks, gluinos...), contain classes of contributions which are enhanced near the partonic threshold
region $\beta \equiv \sqrt{1-4 M^2/\hat{s}} \rightarrow 0$, with  $M=(M_H+M_{H'})/2$ the average 
mass of the two heavy particles and $\hat{s}$ the partonic centre-of-mass energy. 
These corrections arise from soft-gluon radiation off initial- and 
final-state particles ($\sim \alpha_s^n \ln^m \beta$, "threshold logaritms") and
exchange of Coulomb gluons between the two
non-relativistic heavy particles ($\sim (\alpha_s/\beta)^n$, "Coulomb singularities"), and should be 
resummed to all orders in $\alpha_s$ if the hadronic cross section is dominated by the  
partonic threshold region. Resummation of threshold logarithms in the 
Mellin-moment space formalism has been 
discussed, for example, in  \cite{Kidonakis:1997gm, Bonciani:1998vc}, 
and studies of the effects of Coulomb 
resummation in production of top quarks and SUSY particles have been presented in \cite
{Hagiwara:2008df,Kiyo:2008bv,Kulesza:2009kq}. However, the issue of 
factorisation and simultaneous resummation of soft and Coulomb effects in pair production
has been addressed rigorously only recently in \cite{Beneke:2009rj}, employing 
effective-theory techniques similar to ones previously applied in the context of DIS and Drell-Yan
\cite{Becher:2006mr,Becher:2007ty}. 
Here we present the application of results obtained in \cite{Beneke:2009rj} to the process of squark-antisquark production at the LHC, and discuss the numerical relevance of soft and Coulomb resummation for 
theoretical predictions of the total cross section (see also \cite{Beneke:2009nr}).   

\section{Factorisation and RG evolution equations}

In \cite{Beneke:2009rj} it has been shown that near the partonic threshold $\hat{s} \sim 4 M^2$ the 
total partonic cross section for the pair-production process 
factorises, at leading-order in $\beta$
and for particles produced in an $S$-wave state, according to 
\begin{equation} \label{eq:fac} 
\hat\sigma_{pp'}(\beta,\mu_f)
= \sum_{i,i'}H_{ii'}(M,\mu_f)
\;\int d \omega\;
\sum_{R_\alpha}\,J_{R_\alpha}(E-\frac{\omega}{2})\,
W^{R_\alpha}_{ii'}(\omega,\mu_f)\, ,
\label{factform}
\end{equation}
with $E=\sqrt{\hat{s}}-2 M$. Here $H_{i i'}(M,\mu_f)$ denotes a process-dependent hard coefficient 
encoding short-distance effects, decomposed over a basis of 
colour-state operators $\{c^{(i)}_{\{a\}}\}$. The soft function $W^{R_\alpha}_{ii'}(\omega,\mu_f)$ is defined in terms of matrix elements of soft Wilson lines, 
decomposed over the basis
$\{c^{(i)}_{\{a\}}\}$ and projected onto the irreducible colour representations $R_\alpha$ of
the $H H'$ system, $R \otimes R'=\sum_\alpha R_\alpha$.  
The function $J_{R_\alpha}$ describes the internal evolution of the heavy-particle pair $HH'$,  
which is driven by Coulomb-gluon exchange. 
Near the partonic threshold the basis $\{c^{(i)}_{\{a\}}\}$ can be chosen such that the soft function 
$W^{R_\alpha}_{ii'}(\omega,\mu_f)$ is diagonal in colour space to all orders in $\alpha_s$, and the 
colour structure of (\ref{eq:fac}) simplifies significantly \cite{Beneke:2009rj}. 

The resummation of threshold logarithms follows the formalism of \cite{Becher:2006mr}, and is a generalisation of the Drell-Yan case \cite{Becher:2007ty} to arbitrary colour representations $R_\alpha$ of the heavy-particle pair system. 
From results on the IR structure of general massive amplitudes in QCD \cite{Becher:2009kw}
it follows that the diagonal elements of the hard function $H_{i}(M,\mu_f) \equiv H_{ii}(M,\mu_f)$ satisfy 
a renormalisation-group evolution equation given by \cite{Beneke:2009rj} 
\begin{equation} \label{eq:hard}
\frac{d}{d \ln \mu_f} H_i(M,\mu_f)=\left(\gamma_{\mbox{\scriptsize cusp}} (C_r+C_{r'}) \ln \left(\frac{4 M^2}{\mu_f^2}\right)+2 \gamma_i^V \right) \, ,
\end{equation}
where $\gamma_{\mbox{\scriptsize cusp}}$ denotes the cusp anomalous dimension, $C_{r}$ and $C_{r'}$ are the Casimir invariants of the initial-state representations, and $\gamma_i^V$ can be 
written in terms of single-particle anomalous dimensions $\gamma_i^V=\gamma^r+\gamma^{r'}+
\gamma_{H,s}^{R_\alpha}$. From the scale-invariance of the total
hadronic cross section, one can derive an analogous evolution equation for the
soft function $W^{R_\alpha}_i(\omega,\mu_f)$ \cite{Becher:2007ty},
\begin{eqnarray}\label{eq:soft}
   \frac{d}{d \ln \mu_f} W^{R_\alpha}_i(\omega,\mu_f) &=& -2 \left[\left(C_r+C_{r'}\right) \gamma_{\mbox{\scriptsize cusp}} \ln \left(\frac{\omega}{\mu_f}\right)+2 \gamma_{W,i}^{R_\alpha}\right] W^{R_\alpha}_i (\omega,\mu_f) \nonumber \\
   && -2 \left(C_r+C_{r'} \right) \gamma_{\mbox{\scriptsize cusp}}
   \int_0^\omega d \omega' \frac{W^{R_\alpha}_i(\omega',\mu_f)-W_i^{R_\alpha}(\omega,\mu_f)}{\omega-\omega'}  \, ,
  \end{eqnarray}
with the anomalous dimension $\gamma_{W,i}^{R_\alpha}$ given by $\gamma^{R_\alpha}_{W,i}=
\gamma_i^V+\gamma^{\phi,r}+\gamma^{\phi,r'}$, and $\gamma^{\phi,r}$ defined by the large-$x$ 
limit of the Altarelli-Parisi
splitting function
\begin{equation}
P_{p/k}(x,\mu_f)=\left(2 C_r \gamma_{\mbox{\scriptsize cusp}} \frac{1}{[1-x]_+}+
2 \gamma^{\phi,r} \delta(1-x)\right) \delta_{pk}+... \, .
\end{equation}

\section{Momentum-space resummation}

\begin{figure}[t]
\begin{center}
  \includegraphics[width=0.45 \linewidth]{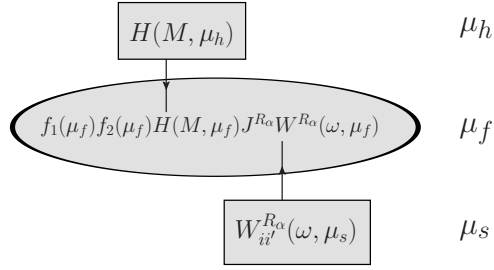}
\end{center}
\caption{Schematic representation of the resummation of threshold logarithms via RGEs.}
\label{fig:RGE}
\end{figure}
In the approach of \cite{Becher:2006mr} the resummation of threshold logarithms is obtained  
directly in momentum space by calculating 
the hard and soft functions at scales $\mu_h$ and $\mu_s$ 
respectively, and evolving them to the common factorisation scale $\mu_f$ using the RG
equations (\ref{eq:hard}) and (\ref{eq:soft}), as schematically depicted in figure \ref{fig:RGE}.   
The hard scale $\mu_h$ and soft scale $\mu_s$ are chosen to minimise radiative corrections
to $H_i(M,\mu_h)$ and $W_i^{R_\alpha}(\omega,\mu_s)$. As anticipated above, the solutions of equations
(\ref{eq:hard}) and (\ref{eq:soft}) are generalisations of the Drell-Yan case discussed in \cite{Becher:2007ty} and
read 
\begin{eqnarray} \label{eq:res}
H^{\mbox{\scriptsize res}}_i(M,\mu_f) &=& \exp[4 S(\mu_h,\mu_f)
  -2 a_i^{V}(\mu_h,\mu_f)] \left(\frac{4 M^2}{\mu_h^2}
  \right)^{-2 a_\Gamma(\mu_h,\mu_f)} H_i(M,\mu_h) \, ,\nonumber\\
 W^{R_\alpha,\mbox{\scriptsize res}}_{i}(\omega,\mu_f) &=& \exp[-4 S(\mu_s,\mu_f)
  +2 a^{R_\alpha}_{W,i}(\mu_s,\mu_f)]  \tilde{s}_{i}^{R_\alpha}(\partial_\eta,\mu_s)
 \frac{1}{\omega} \left(\frac{\omega}{\mu_s}\right)^{2 \eta} \theta(\omega)
 \frac{e^{-2 \gamma_E \eta}}{\Gamma(2 \eta)} \, ,
\end{eqnarray}
where
\begin{eqnarray} \label{eq:exp}
&& S(\nu,\mu) = -(C_r+C_{r'}) \int_{\alpha_s(\nu)}^{\alpha_s(\mu)}d \alpha_s
\frac{ \gamma_{\mbox{\tiny cusp}}(\alpha_s)}{2\beta(\alpha_s)}
\int_{\alpha_s(\nu)}^{\alpha_s}\frac{d \alpha_s^{'}}{\beta(\alpha_s^{'})} \, ,
\nonumber\\
&& a_{\Gamma}(\nu,\mu) =
-(C_r+C_{r'}) \int_{\alpha_s(\nu)}^{\alpha_s(\mu)} d \alpha_s
\frac{\gamma_{\mbox{\tiny cusp}}(\alpha_s)}{2\beta(\alpha_s)} \, ,
 \hspace{0.5 cm}
a^{X}_i (\nu,\mu) =
-\int_{\alpha_s(\nu)}^{\alpha_s(\mu)} d \alpha_s \frac{
\gamma_i^{X}(\alpha_s)}{\beta(\alpha_s)} \, ,
\end{eqnarray}
and $\tilde{s}_{i}^{R_\alpha}(\rho,\mu_s)$ is the Laplace-transform of $W_i^{R_\alpha}(\omega,\mu_s)$ with respect to the variable $s=1/(e^{\gamma_E} \mu e^{\rho/2})$,
\begin{equation}
\tilde{s}_{i}^{R_\alpha}(\rho,\mu_s)=\int_{0_{-}}^\infty d \omega e^{-s \omega} W^{R_\alpha}_i(\omega,\mu_s) \, .
\end{equation}
The one-loop expression for $\tilde{s}_{i}^{R_
\alpha}(\partial_\eta,\mu_s)$ for arbitrary colour representations was computed in
\cite{Beneke:2009rj}. Notice that the resummed hard and soft functions in eq. (\ref{eq:res}) are formally
independent of the choice of $\mu_h$ and $\mu_s$, but when 
$H(M,\mu_h)$ and $\tilde{s}_{i}^{R_\alpha}(\partial_\eta,\mu_s)$ are
truncated at a finite perturbative order a dependence on the hard and soft scales is introduced, 
which is however of higher order in $\alpha_s$. 

The resummation of the velocity-enhanced terms $(\alpha_s/\beta)^n$ associated with 
Coulomb-gluon exchange is obtained by relating the potential function $J_{R_\alpha}$ to the zero-distance Coulomb Green function, $J_{R_\alpha}=2 \mbox{Im} G^{(0)}_{C,R_\alpha}(0,0;E)$. 
Using the representation provided in \cite{Beneke:1999zr}, this gives
\begin{eqnarray} \label{eq:J} 
J_{R_\alpha}(E) &=& -\frac{(2m_{\mbox{\scriptsize red}})^2}{2 \pi} \mbox{Im} \Bigg\{\!
    \sqrt{-\frac{E}{2m_{\mbox{\scriptsize red}}}}
    + \alpha_s(-D_{R_\alpha}) \bigg[
      \frac{1}{2}\ln \bigg(\! -\!\frac{8\,m_{\mbox{\scriptsize red}}E}{\mu_f^2}\!\bigg)\! \nonumber \\
   &&  
   -\frac{1}{2}  +\gamma_E
    +\psi\bigg(1-\frac{\alpha_s(-D_{R_\alpha})}{
      2\sqrt{-E/ (2m_{\mbox{\scriptsize red}})}}\!\bigg)\!\bigg]\!\Bigg\} \,  ,
\end{eqnarray}   
with $D_{R_\alpha}=(C_{R_\alpha}-C_R-C_{R'})/2$ and $m_{\mbox{\scriptsize red}}=M_H M_{H'}/(M_H+M_{H'})$. The second term on the first line of (\ref{eq:J}) encodes the contribution to the
cross section from a single Coulomb-exchange diagram, while the second line accounts for the contribution of two or more gluons. Beside modifying the shape of the cross section above threshold ($E>0$), the resummation of Coulomb
effects leads to the appearance of heavy-particle bound states below threshold for 
an attractive Coulomb potential ($D_{R_\alpha}<0$). In the limit
of vanishing width, $\Gamma_H=\Gamma_{H'}=0$, and for $E<0$, 
$J_{R_\alpha}$ reduces to
\begin{equation}\label{eq:BS}
J_{R_\alpha}(E) =  2 \sum_{n=1}^{\infty}  \left(\frac{m_{\tilde{q}} \alpha_s (-D_{R_\alpha})}{2 n}\right)^3 
 \delta \left(E+m_{\tilde{q}} \left(\frac{\alpha_s (-D_{R_\alpha})}{2 n}\right)^2\right) \theta(-D_{R_\alpha}) \, . 
\end{equation}

\section{Threshold resummation for squark-antisquark production}

We now apply the results presented in the previous section to the NLL soft resummation
and Coulomb resummation of the squark-antisquark total production cross section. 
The resummed cross section for the partonic channel $pp'$ is
\begin{equation} \label{eq:res_cross}
\hat\sigma^{\mbox{\scriptsize res}}_{p p'}(\hat s,\mu_f)= \sum_{i} H^{\mbox{\footnotesize NLL}}_{i}(m_{\tilde{q}},\mu_f)
    \int d \omega \;
    \sum_{R_{\alpha}} J_{R_{\alpha}}(E-\frac{\omega}{2})\;
     W^{R_{\alpha},\mbox{\footnotesize NLL}}_{i}(\omega,\mu_f) \, .
\end{equation}  
The NLL resummed hard and soft functions in (\ref{eq:res_cross}) are given by the expressions in eq. (\ref{eq:res}),
where the cusp anomalous dimension $\gamma_{\mbox{\scriptsize cusp}}$ is included at the 
two-loop accuracy, the anomalous dimensions $\gamma^V_i$, $\gamma_{W,i}^{R_\alpha}$ at 
one-loop accuracy, and $H_i(M,\mu_h)$ and $\tilde{s}_{i}^{R_\alpha}(\partial_\eta,\mu_s)$ are 
replaced by their tree-level expressions $H_i^{(0)}(M,\mu_h)$ and
 $\tilde{s}_{i}^{R_\alpha (0)}(\partial_\eta,\mu_s) = 1$ (for more details on the reorganised
 perturbative expansion in presence of soft and Coulomb resummation we refer to \cite{Beneke:2009rj}) . The resummed cross section (\ref{eq:res_cross}) is then matched onto the full fixed-order NLO result   
\begin{equation} \label{eq:match}
 \hat\sigma^{\mbox{\scriptsize match}}_{p p'}(\hat s,\mu_f)=
   \left[\hat\sigma^{\mbox{\scriptsize res}}_{p p'}(\hat s,\mu_f)-
   \hat\sigma^{\mbox{\scriptsize res}}_{p p'}(\hat s,\mu_f)|_{\mbox{\scriptsize NLO}}\right]
   +\hat\sigma^{\mbox{ \scriptsize NLO}}_{p p'}(\hat s,\mu_f) \, ,
\end{equation}
using the parametrisation of the NLO cross section extracted in 
\cite{Langenfeld:2009eg} from numerical
results obtained with Prospino \cite{Beenakker:1996ch}. 
Our final prediction for the hadronic cross section
is given by the convolution of (\ref{eq:match}) with the MSTW2008 NLO PDFs.
The default value for the soft scale, denoted with $\tilde{\mu}_s$, is chosen such that the 
one-loop soft correction to the hadronic cross section is minimised. This is 
analogous to the procedure introduced in \cite{Becher:2007ty}, and yields 
$\bar{\mu}_s=123-455 \, \mbox{GeV}$ for the squark mass range 
$m_{\tilde{q}}=200 \, \mbox{GeV}-2 \, \mbox{TeV}$. Furthermore, we evaluate
 the running coupling in $J_{R_\alpha}$ at the scale 
$\mu_C = \mbox{Max}\{2 m_{\tilde{q}} \beta, {C_F m_{\tilde{q}} \alpha_s(\mu_C)}\}$. 
This choice is motivated by the characteristic virtuality of a
Coulomb gluon, $q^2 \sim m_{\tilde{q}}^2 \beta^2$, and by the inverse Bohr radius of
the first $HH'$ bound state, $1/r_B=C_F m_{\tilde{q}} \alpha_s/2$.
Finally, we identify the hard and factorisation scale, $\mu_h=\mu_f$,
and choose the latter of order $m_{\tilde{q}}$. This means that the
resummed hard function in (\ref{eq:res_cross}) reduces to its tree-level
expression, 
since the functions (\ref{eq:exp}) vanish
for this particular scale choice. The function $H^{(0)}_{i}(m_{\tilde{q}},\mu_f)$ is extracted
from the expansion at threshold of the tree-level cross section $\sigma_{i}^{(0)}=
\frac{m_{\tilde{q}}}{2 \pi} \beta H^{(0)}_ {i}+O(\beta^3)$.  

\begin{wrapfigure}{r}{8 cm}
\includegraphics[width=\linewidth]{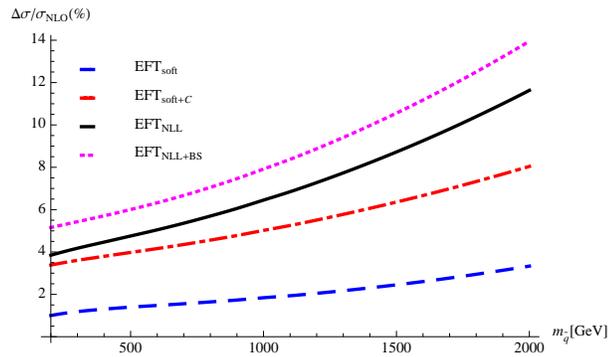}
\caption{Corrections to the NLO squark-antisquark cross section from soft-Coulomb
resummation as functions of the squark mass $m_{\tilde{q}}$.}
\label{fig:NLL}
\end{wrapfigure}
The numerical effects of soft-Coulomb resummation on the total hadronic cross section,
normalised to the full NLO result, are 
shown in figure \ref{fig:NLL}, for $\sqrt{s}=14 \, \mbox{TeV}$, degenerate squark masses,
$m_{\tilde{g}}/m_{\tilde{q}}=1.25$ and $\mu_f=m_{\tilde{q}}$.  
The four curves represent different contributions
contained in the resummed cross section (\ref{eq:res_cross}). The lowest curve ("$\mbox{EFT}_{\mbox{\scriptsize soft}}$", dashed blue) shows the effect of NLL soft-gluon resummation without any Coulomb effect, corresponding to setting $J_{R_\alpha}$ to its tree-level value in
(\ref{eq:res_cross}).
"$\mbox{EFT}_{\mbox{\scriptsize soft+C}}$" (dot-dashed red) includes, on top of soft 
effects, Coulomb resummation, but no interference of the two. This is partially included in the
third curve ("$\mbox{EFT}_{\mbox{\scriptsize NLL}}$", solid black) convoluting the one-loop Coulomb contribution (second
term in (\ref{eq:J})) with the full NLL soft function (see also \cite{Beneke:2009nr}). Finally, the 
contribution to the cross section from bound states below threshold is added in the last curve ("$\mbox{EFT}_{\mbox{\scriptsize NLL+BS}}$", dotted magenta). The pure soft resummation is in qualitative agreement with the
Mellin-space results quoted in \cite{Kulesza:2009kq}, while the choice of a running scale
in the Coulomb corrections leads to much larger effects than in \cite{Kulesza:2009kq}, where
$\alpha_s$ at the scale $\mu_f$ was used. Also, the soft-Coulomb interference and
bound-state contributions shown in figure \ref{fig:NLL} turn out to be comparable in size
to pure soft and Coulomb effects, and are therefore necessary for precise theoretical predictions. 

\begin{figure}[t]
\includegraphics[width=0.47 \linewidth]{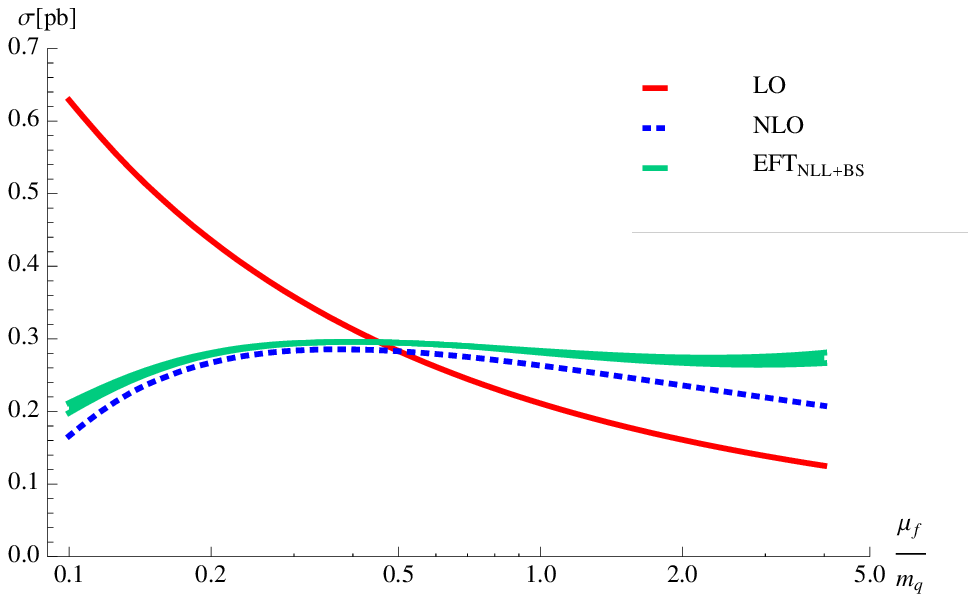}
\includegraphics[width=0.47 \linewidth]{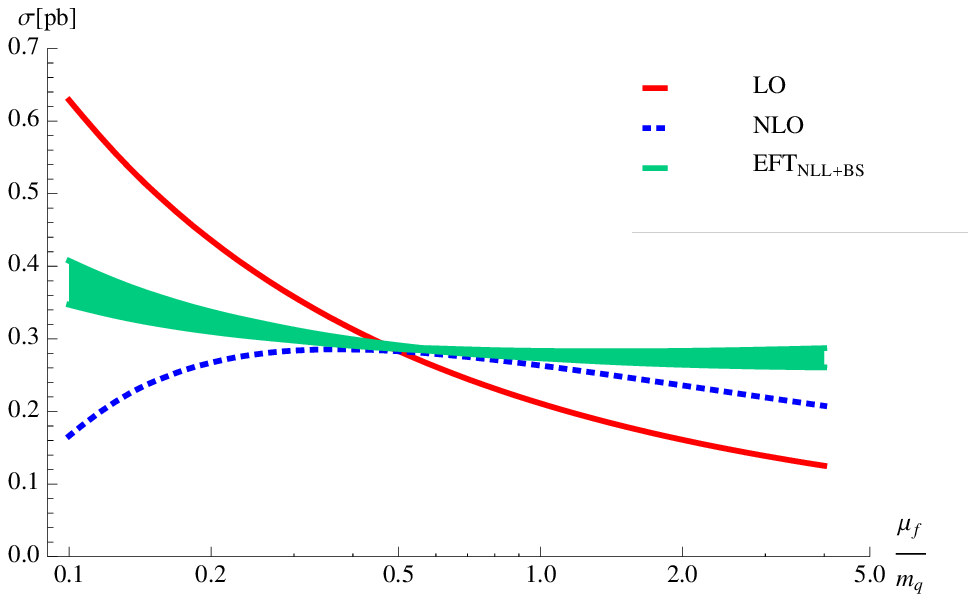}
\caption{Scale dependence of the LO (solid red), NLO (dotted blue) and resummed cross section (green band) for different choices of $\mu_s$ and $\mu_h$. See the text for explanation.}
\label{fig:scale}
\end{figure}

The reduced scale dependence of the resummed cross section is presented in figure \ref{fig:scale}. 
The left plot shows the LO, NLO and resummed result as functions of the factorisation 
scale $\mu_f$, for $m_{\tilde{q}}=1 \, \mbox{TeV}$ and $\mu_h=\mu_f$. The
green band represents the uncertainty associated with the choice of the soft scale, which 
varies in the interval $\bar{\mu}_s/2<\mu_s<2 \bar{\mu}_s$. The 
scale dependence is clearly reduced for large $\mu_f$, but only
mildly improved at small values of the scale, where $\ln (\mu_s/\mu_f)=O(1)$ and the effect of 
soft resummation is negligible. It can be argued that 
in this region ($\mu_f \lesssim 0.2 \, m_{\tilde{q}}$) the 
identification of the hard scale with the factorisation scale is not justified, 
and the two scales should be kept separate.
This is shown in the right plot in figure \ref{fig:scale}. In this case the green band is obtained
by varying $\mu_h$ and $\mu_s$, independently from the factorisation scale, 
in the intervals $m_{\tilde{q}}<\mu_h<4 m_{\tilde{q}}$ and $\bar{\mu}_s/2<\mu_s<2 \bar{\mu}_s$,
respectively.
While the cross section is not significantly modified with respect to the left plot at large scales, in the region $\mu_f \ll m_{\tilde{q}}$ the appearance of large logarithms of $\mu_h/\mu_f$, and their resummation in $H^{\mbox{\footnotesize NLL}}_{i}(m_{\tilde{q}},\mu_f)$, leads to  sizeable corrections to the cross section.

\end{document}